# MyI-Net: Fully Automatic Detection and Quantification of Myocardial Infarction from Cardiovascular MRI Images

**Shuihua Wang [1], Senior member, IEEE, Ahmed M.S.E.K Abdelaty[2], Kelly Parke[2], J Ranjit Arnold[2], Gerry P McCann[2] and Ivan Y Tyukin[3], Member, IEEE**

[1] Department of Cardiovascular Sciences, University of Leicester, the NIHR Leicester Biomedical Research Centre, Glenfield Hospital, Leicester, and School of Computing and Mathematical Sciences, University of Leicester, United Kingdom; sw546@le.ac.uk

[2] Department of Cardiovascular Sciences, University of Leicester, and the NIHR Leicester Biomedical Research Centre, Glenfield Hospital, Leicester, United Kingdom. amseka2@le.ac.uk, kelly.parke@uhl-tr.nhs.uk, jra14@le.ac.uk; gpm12@le.ac.uk

[3] Department of Mathematics, King's College London, United Kingdom; Norwegian University of Science and Technology (NTNU), Norway; Saint-Petersburg State Electrotechnical University, Russia, and Lobachevsky University, Russia; ivan.tyukin@kcl.ac.uk

\* Correspondence: gpm12@le.ac.uk, ivan.tyukin@kcl.ac.uk;

**Abstract:** A ''heart attack'' or myocardial infarction (MI), occurs when an artery supplying blood to the heart is abruptly occluded. The "gold standard" method for imaging MI is Cardiovascular Magnetic Resonance Imaging (MRI), with intravenously administered gadolinium-based contrast (late gadolinium enhancement [LGE]). However, no "gold standard" fully automated method for the quantification of MI exists. In this work, we propose an end-to-end fully automatic system (MyI-Net) for the detection and quantification of MI in MRI images. This has the potential to reduce the uncertainty due to the technical variability across labs and inherent problems of the data and labels. Our system consists of four processing stages designed to maintain the flow of information across scales. First, features from raw MRI images are generated using feature extractors built on ResNet and MoblieNet architectures. This is followed by the Atrous Spatial Pyramid Pooling (ASPP) to produce spatial information at different scales to preserve more image context. High-level features from ASPP and initial low-level features are concatenated at the third stage and then passed to the fourth stage where spatial information is recovered via up-sampling to produce final image segmentation output into: i) background, ii) heart muscle, iii) blood and iv) scar areas. Experiments showed that the model named as MI-ResNet50-AC provides the best global accuracy (97.38%), mean accuracy (86.01%), weighted Intersection over Union (IoU) of 96.47%, and bfscore of 64.46% for the global segmentation. However, in detecting only scar tissue, a smaller model, MI-ResNet18-AC, showed higher accuracy (74.41%) than MI-ResNet50-AC (64.29%). New models were compared with state-of-art models and manual quantification. Our models showed favorable performance in global segmentation and scar tissue detection relative to state-of-the-art work, including a four-fold better performance in matching scar pixels to contours produced by clinicians.

## 1. Introduction

Myocardial infarction (MI), commonly referred to as a 'heart attack' occurs when an artery supplying blood to the heart is abruptly occluded. It is caused by the rupture of an atherosclerotic plaque in the wall of the artery, triggering the clotting cascade and leading to vessel occlusion. This may result in severe damage to the heart muscle which may be irreversible (scar). The extent of scarring following more severe heart attacks (ST segment elevation MI, or STEMI) may drive enlargement of the heart, and is associated with worse prognosis (increased risk of death and subsequent heart failure) [1, 2]. According to a report from the British Heart Foundation (BHF) in 2020, MI accounts for approximately 100,000 hospital admission annually. It is estimated that there are 1.4 million individuals alive in the UK today who have survived an MI (1 million men and 380,000 women) [3].

Cardiovascular magnetic resonance Imaging (MRI) provides accurate non-invasive diagnosis of MI. The late gadolinium enhancement (or LGE) technique [4, 5], uses gadolinium-based contrast agent and specified magnetic resonance pulse sequences to provide a reproducible method for identifying and quantifying MI. LGE-CMR is recognized as the "gold standard" non-invasive method for visualizing and diagnosing MI, and also provides vital prognostic information following MI. A number of methods are available for the quantitative assessment of MI size (from LGE images), including visual assessment, manual planimetry, and semi-quantitative methods (such as full width at half maximum [FWHM]) [6, 7]. However, to date, there is no "gold standard" fully automatic method for MI detection and quantification.

In the past decades, several groups of researchers have been working to develop either semiautomatic or fully automatic methods for the detection and quantification of MI from MRI scans. For example, Eitel, et al. [8] proposed a standard deviation (SD) method for the quantification of the salvaged myocardium area extent after reperfusion. Amado, et al. [9] used the FWHM criterion to confirm that MI can be sized accurately up to 30 minutes after contrast administration. Flett, et al. [10] compared seven quantification methods including manual quantification, 2, 3, 4, 5, or 6 SDs above remote myocardium, and the full FWHM method. They confirmed that FWHM methods provide the closest result to manual quantification and has the highest reproducibility. Hsu, et al. [11] measured the MI size of 11 dogs based on the automated feature analysis and combined thresholding (FACT). Comparison of the proposed FACT algorithm with FWHM, intensity thresholding and human manual contouring confirmed that human contouring may overestimate MI size and the higher accuracy can be obtained from FACT than intensity thresholding. Tong, et al. [12] proposed the current interleaved attention network (RIANet) for the cardiac MRI segmentation based on ACDC 2017. Shan, et al. [13] proposed the segmentation method based on spatiotemporal generative adversarial learning without contrast agents. Xu, et al. [14] proposed long short-term memory recurrent neural network (LSTM-RNN) for MI detection without contrast agents. Héloïse Bleton [15] proposed left ventricular infarct location based on Neighbourhood Approximation Forests (NAF) and compared with the stack autoencoder method based on 4D cardiac sequences. Fahmy, et al. [16] developed a UNet DCNN model for automatic cardiac MI quantification with stratified random sampling. Bernard, et al. [17]'s review reported that for the ACDC2017 challenge of cardiac MRI assessment, many researchers proposed the using of UNet [18] for the segmentation of myocardium, right ventricle and left ventricle. Fahmy, et al. [19] also proposed using the UNet method for the MI segmentation based on data collected from patients with and without MI.

Although significant progress has already been made in assisting clinical experts to quantify the size of MI in affected patients, major hurdles still remain in this vitally important area. For example, manual tracing of contours is subjective and prone to low reproducibility with high intra- and interobserver variability, as well as being labor-intensive, with associated costs. Existing semi-automatic methods to localize MI are affected by biases introduced through tracing of the left ventricle (LV). All those challenges introduce significant uncertainty when detecting and quantifying the scar from Cardiovascular MRI Images

In this paper, we propose a new system, named MyI-Net, to achieve end to end, fully automatic MI detection and quantification to overcome above challenges. In order to optimize performance, we propose a new class of appropriately engineered deep leaning models. These models combine initial feature extraction (realized through ResNet and MobileNet – based models) followed by the Atrous Spatial Pyramid Pooling (ASPP) to adjust the receptive field to preserve more image context. New feature maps are generated via fusing high-level features from ASPP and low-level features from one specific layer of corresponding networks. Finally, the segmentation result is obtained via up-sampling to eventually recover the spatial information by an add-on module.

In order to deal with the other source of uncertainty, the issue of inherently unbalanced datasets (the number of pixels corresponding to scarred tissue in an image is always considerably smaller than that of the pixels corresponding to muscle, background, or blood pool) while fully using all the data, we use an appropriately constructed weight matrix. As training datasets are always limited, and in order to increase robustness, we propose three different augmentation methods integrated in this model to make a diversified dataset for training. New models as well as their state-of-the-art counterparts which were used as baseline comparisons were trained and validated on 1822 unique MRI images collected in our lab from research patients with MI.

The rest of this paper is organized as follows: Section 2 provides a detailed account of materials, procedures of the data collection and the demographics of data. Section 3 presents our proposed methods including construction of the weight matrix, details of data augmentation and detailed information regarding performance metrics. Section 4 presents results of our experiments including time cost analysis, segmentation analysis and comparison with state of the art work as well as with manual segmentation produced by a human expert. Section 5 concludes and illustrates our proposed method, its limitations and future research.

## 2. Materials

The data was collected using Cardiovascular magnetic resonance (CMR) imaging. With gadolinium-based contrast agents and appropriate pulse sequences, CMR can provide clear differentiation between the infarcted and normal myocardium. To obtain LGE images, the patient is typically scanned 10-20 minutes after the intravenous administration of standardized, weight-adjusted dose of gadolinium-based contrast agent.

The data came from a variety of MRI scanners: for data collected from Siemens 1.5T scanners, the sequence parameters are as follows: slice thickness was 10mm, repetition time was 900ms, echo time was 4.91ms, flip angle 30o and Acquisition matrix – 256/154. For Philips 1.5T, slice thickness was 10mm, repetition time was 4.87ms, echo time was 1.87ms, acquisition matrix – 256/256. For Siemens 3T Skyra, slice thickness was 8mm with 2mm gap, repetition time was 43.29ms, echo Time 1.46ms and acquisition matrix – 256/208. The data from different vendors and field strengths enables generalization of model results. The data is collected from patients with MI whose demographic data are shown in Table 1.

**Table 1**. Demographic data

| Variables | Values | Number | Rate |
|---|---|---|---|
| Gender | Male | 172 | 57.14% |
| | Female | 129 | 42.86% |
| Age at MRI | 90-99 | 3 | 1% |
| | 80-89 | 11 | 3.65% |
| | 70-79 | 44 | 14.62% |
| | 60-69 | 77 | 25.58% |
| | 50-59 | 74 | 24.58% |
| | 49-49 | 56 | 18.60% |
| | 30-39 | 35 | 11.63% |
| | 20-29 | 1 | 0.33% |

### 3. Automated segmentation of myocardial infarction: myocardial infarction-Net (MyI-NET)

As reported by the Association of American Medical Colleges (AAMC), in the US there will be an urgent shortage of physicians (approximately 122,000 by 2032) while the nation's population is still growing and aging [20]. A similar situation is expected to emerge in the UK as only 2% of radiology departments have the ability to fulfil their imaging interpretation tasks within work hours as reported by the Royal College of Radiologists (RCR) entitled 'Clinical Radiology U.K. Workforce Census Report 2018' [21]. Meanwhile, the report also highlighted that only 2% trusts and health boards in the UK have adequate interventional radiologists to provide for urgent procedures. Therefore, an automatic image interpretation service is urgently needed.

Deep learning has demonstrated great potential in biomedical data analysis with its powerful and advanced learning abilities [22-24]. For example, Nam, et al. [**Error! Bookmark not defined.**] reported in 2018 that their proposed algorithm for malignant pulmonary nodules based on deep learning outperforms the radiologist in radiograph classification. Our pilot work based on CNN [26] also illustrated that deep learning can be used for automatic detection of MI. Here we make another step forward to improve the performance of MI detection based on machine learning, by proposing a new class of models: MyI-Net. Details of the proposed new class are provided below.

At the core of this new model class is the proposal to exploit a wealth of deep learning architectures whose efficiency has already been demonstrated in image processing applications. We will use these models as a part of the feature extraction process. Feature extraction is then combined with the Atrous spatial pooling – ASPP. The latter generates multiple receptive fields enabling us to catch information at different spatial scales in a balanced way. This is followed by an add-on module for the spatial information recovery. The process, applied to MRI MI segmentation, is illustrated with a diagram shown in Figure 1. In Figure 1, $X = (x_1, x_2, ..., x_n)$ stands for the low-level features that can be extracted from the specific i-th layer of the base feature extraction network. Core outputs of the backbone deep learning model which have been used in the initial processing pipeline are referred to as the high-level features.

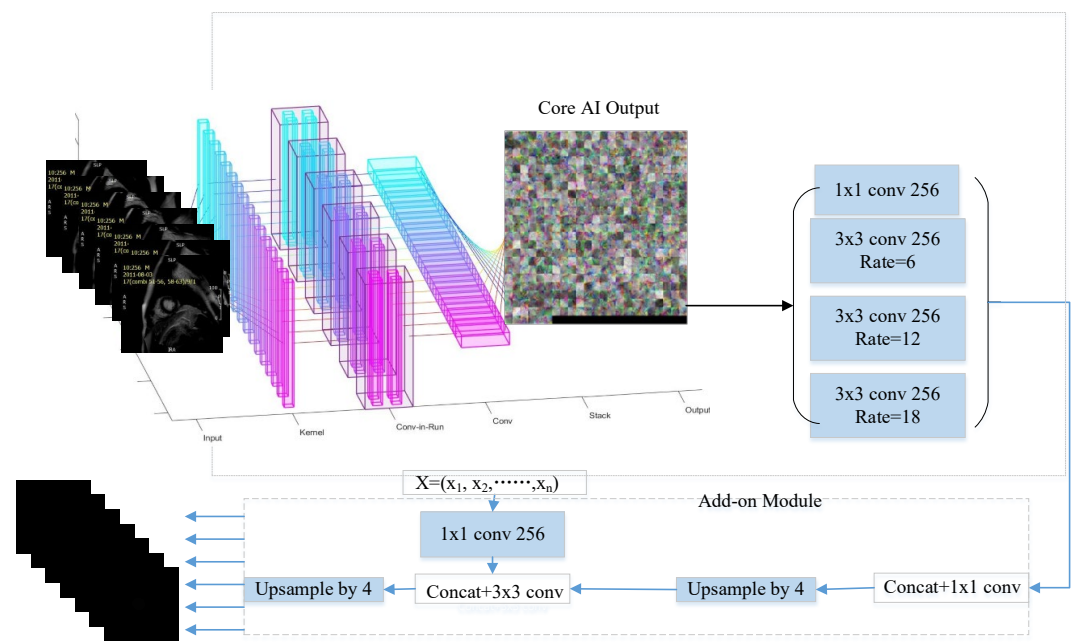

**Figure 1.** Flowchart of the proposed model.

### 3.1. Feature extraction by MI-ResNet

Feature extraction is based on deep CNN networks (see e.g. [25, 27, 28]). Figure 2 shows an example flowchart of relevant processes in a conventional CNN. As is shown in Figure 2, all layers, including convolutional layers, ReLu layers, pooling layers, are cascaded gradually. However, such simple and uniformly cascaded structures have severe technical drawbacks. Particularly, it may be hard to train deep conventional CNNs in practice due to the well-known problems of either gradient exploding or gradient vanishing. To circumvent this issue, here we adopt the ResNet model of CNN proposed by He, et al. [29] as the basic backbone model for feature extraction. We call this backbone model MI-ResNet. In contrast to conventional CNNs, ResNet provides a structure with short-cut connections by skipping one or more weight layers as shown in Figure 3.

Mathematically, the structure of CNN and ResNet processing blocks can be expressed as:

$$\text{CNN: } R_l = H_l \quad (1)$$

$$\text{ResNet: } R_l = H_l + x_{l-1} \quad (2)$$

in which, $x_{l-1}$ stands for the output from the previous layer, H_l is the output of the l-th layer in the conventional CNN's counterpart, and $R_l$ is the output of a ResNet constructed from the original CNN by adding a short-cut connection (residual information).

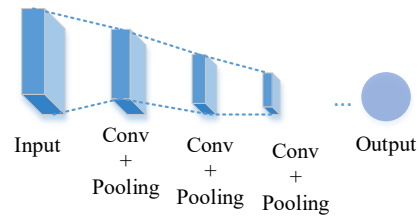

**Figure 2.** Structure of conventional CNN.

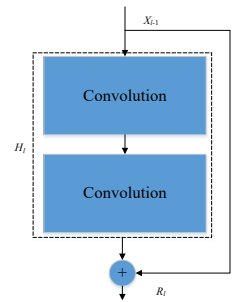

**Figure 3.** Short-cut Structure of the ResNet Block.

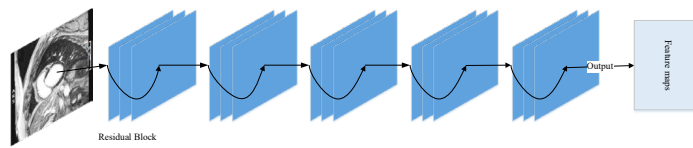

**Figure 4.** Structure of MI-ResNet.

### 3.2. Feature extraction by MI-MobileNet

ResNet architectures mainly focus on improving the accuracy of the deep network and ignore computation costs. Therefore, we consider MobileNetV2 as another potentially relevant backbone for our proposed MyI-Net. We call such architectures MI-MobileNet feature extractors. MobileNetV2 was proposed by Sandler, et al. [30], a research group in Google. Before the advent of MobileV2, MobileNet was first introduced by Howard, et al. [31] also in Google with the idea of depthwise separable convolution (DSC), which can dramatically reduce the model size and complexity. DSC can be described as by two components: depthwise convolution (DC) and pointwise convolution. DC applies a single filter to each input channel and the pointwise convolution applies 1x1 filters to create a linear combination of the output of DC layers. There also are batch normalization layers and ReLu layers to follow both the DC layer and pointwise convolution layer. The structure of DSC is shown in Figure 5. As the DC in MobileNet used the 3x3 filter, we therefore used the 3x3 filter in Figure 5 to show the difference between the structures of the standard convolution and DSC.

Though MobileNet is rather small and computationally cost efficient, to make it more flexible in practical applications with the requirement of faster running and smaller structure, MobileNet utilizes the idea of the so-called width and resolution multipliers. Width multiplier makes the network uniformly thin at each layer, and the resolution multiplier is applied to the input image to further reduce the computation cost. The detail is introduced as follows:

Suppose that the width multiplier is $\varepsilon$, then for a given layer with the number of input channels $D_I$, the number of input channels become $\varepsilon D_I$. Likewise, if $D_O$ is the number of output channels then this layer's number of output channels will become $\varepsilon D_O$. Therefore, the computation cost of one DSC can be reduced to:

$$\text{Cost}_{\text{DSC}} = \varepsilon D_I * D_F * D_F + \varepsilon D_I * \varepsilon D_O * D_F * D_F \quad (3)$$

where, $D_F$ stands for the spatial width and height of a square input feature map, and $\varepsilon$ is taken in the interval $(0\ \ 1]$. If we set the resolution multiplier equal to $\delta \in (0\ \ 1]$ as well, then the computation cost can be described as:

$$\text{Cost}_{\text{DSC}} = \varepsilon D_I * \delta D_F * \delta D_F + \varepsilon D_I * \varepsilon D_O * \delta D_F * \delta D_F \quad (4)$$

In summary, MobileNet replaces the standard convolution by DSC with batch normalization and ReLU excluding the first full convolution layer. In addition, it enables further reduction of the overall computation costs by using appropriate width resolution multipliers.

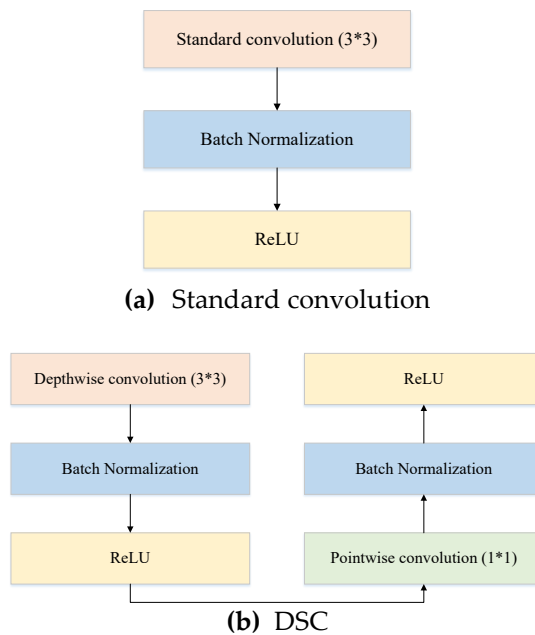

**(a)** Standard convolution

**(b)** DSC

**Figure 5.** The structures of standard convolution and DSC.

MobileNetV2 architecture is a descendant from the base MobileNet in which further processing operations are added, namely bottlenecks. For the bottlenecks, there are two types of blocks; residual block and down-sizing block as shown in Figure 6 (a) and Figure 6 (b) respectively. As is shown in Figure 6, a bottleneck in MobileNetV2 is characterized by that the first layer is the 1x1 convolution layer followed by ReLU6, the second layer is the DC layer and the final layer is 1x1 convolution layer without any non-linear operation. In the residual block, the input of the corresponding block is combined with the output of final 1x1 convolution layer. The whole structure of MobileNetV2 can be found in [30]. For the feature extraction, MI-MobileNet takes the raw LGE-CMR image and generates corresponding features following the MobileNetV2 processing flowchart.

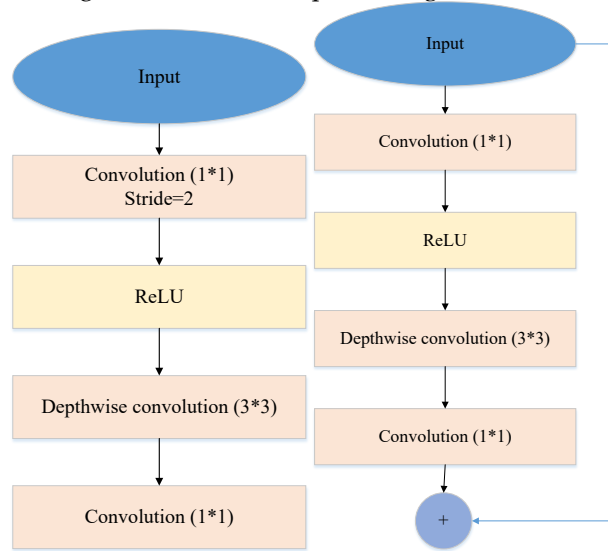

(**a**) Residual block (Stride=1) (**b**) Down-sizing block (Stride=2)

**Figure 6.** Residual block and Down-sizing block in MobileNetV2

### 3.3. Atrous Spatial Pyramid pooling

In the conventional convolution neural network, we can obtain more low-level and high-level features while the network goes deeper and wider. However, the problem is that this standard approach produces a relatively limited number of spatially local convolutional features. The latter, however may contain crucial information for semantic segmentation. For the semantic segmentation, conventional approaches therefore employ different methods to increase spatially relevant information content, like stacking more layers and up-sampling. Theoretically, the amount of spatially relevant information can be increased through a broader spectrum of convolutional filters used in the network: from small to large. The size of these filters is sometimes referred to as a receptive field. Thus, the overall receptive field sizes can be increased if we stack more layers. However, not all information in the receptive fields is equally effective or useful. Likewise, the up-sampling increases the receptive field but at the same time may negatively affect our capability to extract useful information about local context. In order to keep the context information, which is important, and decrease the ambiguity caused by local areas while maintaining the number of parameters in the receptive fields constant, Atrous convolution, also named as dilated convolution, was proposed [32]. Atrous convolution is implemented via assigning zero values in the relevant weights of the filter. Formally, it can be expressed as:

$$h(i) = \sum_{n=1}^{N} f(i + kn).w(n) \quad (5)$$

where, $k$ stands for the dilation rate. When $k = 1$, it reverts to a conventional convolution. $w(n)$ represents the filter with size $n$, $f(i)$ is the input and $h(i)$ is the output of the Atrous convolution. Figure 7 shows examples of the Atrous convolution with rate k=1, 2, 3. When k=2, 3, we obtain feature maps with larger receptive field as shown in Figure 7 (b) and Figure 7 (c).

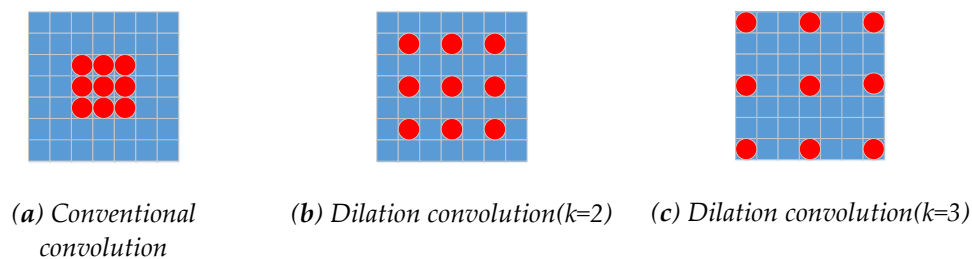

(**a**) *Conventional convolution*    (**b**) *Dilation convolution(k=2)*    (**c**) *Dilation convolution(k=3)*

**Figure 7.** Atrous convolution (the red dot means non-zero)

In this paper, we propose using the Atrous Spatial Pyramid pooling (ASPP) method as an extra module cascaded to the feature extraction network as shown in *Figure 1*. This extra module enables us to adjust and maintain constant size (weights-wise) of receptive fields across scales in the network. ASPP feature maps are generated via a 1x1 convolutions and three Atrous convolution with rate *k, 2k and 3k*. In the models we generated in this work, the value of *k* was set to 6. Their outputs are then fused together to form new feature maps.

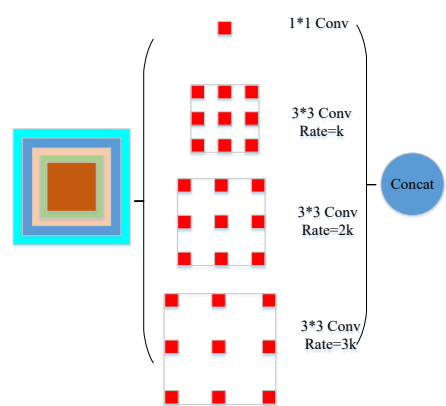

**Figure 8.** Structure of Atrous spatial pooling with feature extraction network

### 3.4. Weight matrix

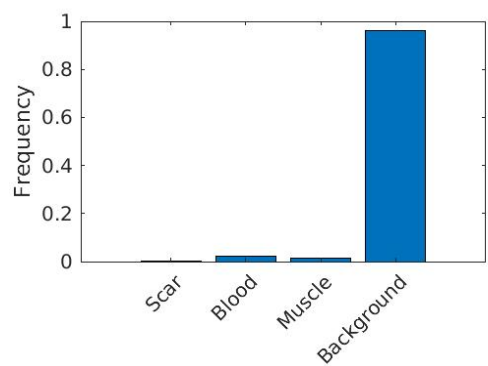

**Figure 9.** Statistical data of each element in the raw images.

Figure 9 shows class frequencies of data in our dataset. As we can see from this figure the dataset is severely imbalanced. Imbalanced datasets, if processed without due care, could produce models that are biased towards the most common category.

In order to avoid this problem and at the same time to utilize our data fully, we employ an appropriately chosen weight matrix to balance the contribution of data from different-sized classes whilst training the model.

Weight matrix assigns appropriate weights to each training sample when a training algorithm computes and subsequently uses a given loss function. In this work, the highest weight is assigned to data associated with the scar tissue, and smallest weight is assigned to data samples representing background pixels. Mathematically, the weight matrix we used is defined as follows:

$$F_i = \frac{N_i}{\sum_{i=1}^{n} N_i} \quad , i = 1, 2, \dots, n \quad (6)$$

$$W_i = \frac{Median\{F_i\}}{F_i} \quad (7)$$

There, $N_i$ represents the number of pixels in each class of the dataset, $n$ represents the total number of categories/classes, $F_i$ stands for the frequency, $i$ represents indices, and $W_i$ is the weight of each category/class.

### 3.5. Data augmentation

As the dataset to train, test, and validate our models was very limited (particularly for scar tissue), data augmentation was utilized to produce a more diversified dataset. The data augmentation methods used in this work include geometric transformations, such as rotation (from 0 to 360 degrees at random) and random scaling with scaling factors from 0.9 to 1.1 for the training dataset in Figure 10.

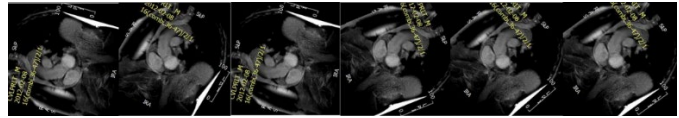

*(a) Random rotation within [0 360]*

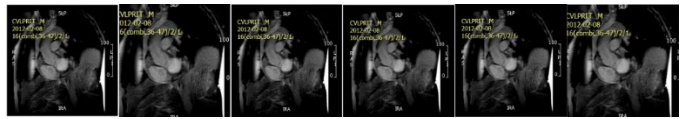

*(b) Random scaling within [0.9 1.1]*

**Figure 10.** Examples of data augmentation

### 3.6. Performance metrics

In order to validate the proposed methods, we employed different performance metrics, including accuracy, bfscore, IoU, and per-image score.

Accuracy: accuracy at the pixel level is defined as the percentage of correctly identified pixels for each category, which is used by most semantic segmentation. Suppose the confusion matrix P, which stands for all the prediction results for the whole dataset is:

$$P_{ab} = \sum_{I \in D} \left| \{ z \in I \ and \ S_g^I(z) = a \ and \ S_p^I(z) = b \} \right| \quad (8)$$

where $z$ stands for each pixel in the image $I$ and $S_g^I(z)$ stands for the ground truth and $S_p^I(z)$ is the prediction result for $z$. $P_{ab}$ is the total number of pixels with label $a$ and prediction output $b$. If we have $n$ categories, then we can get $M_a = \sum_{b=1}^{n} P_{ab}$ as the total number of pixels with label $a$. $H_b = \sum_a P_{ab}$ as the number of pixels predicted as $b$. Then the global accuracy can be expressed as:

$$gAcc = \frac{\sum_{a=1}^{n} P_{aa}}{\sum_{a=1}^{n} M_a} \quad (9)$$

A category accuracy is the total fraction of correctly detected pixels in that category. The global accuracy $gAcc$ is the fraction of all correctly detected pixels regardless of the category information, which can provide a quick and inexpensive measure of the segmentation algorithm. The mean accuracy is the average category accuracies:

$$aAcc = \frac{1}{n}\sum_{a=1}^{n}\frac{P_{aa}}{M_a} \quad (10)$$

Bfscore provides the information of how well the predicted boundary aligns with the ground truth boundary. As the contour quality contributes significantly to the segmentation result, therefore, in this research, we proposed to use the bfscore as one measure, which is mathematically expressed as the harmonic mean of the recall $R_o$ and precision $P_o$ to determine whether the predicted boundary matches to the ground truth boundary with a distance error tolerance ∂. The detail description is as follows:

Let $B_g^o$ be the boundary of the binary ground truth segmentation map for a specific class $o$ with $S_g^o(z) = [\![S_g(z) == o\ ]\!]$ and $[\![z]\!]$ be the Iverson bracket notation (2):

$$[\![z]\!] = \begin{cases} 1 & if\ z = true \\ 0 & otherwise \end{cases} \quad (11)$$

Let $B_p^o$ be the predicted binary contour map for the segmentation result $S_p^o$. Then, with a distance error tolerance $\partial$, precision and recall for each class are defined as

$$P_o = \frac{1}{|B_p|}\sum_{z \in B_p^o}[\![d(z, B_g^o) < \partial]\!] \quad (12)$$

$$R_o = \frac{1}{|B_g|}\sum_{z \in B_g^o}[\![d(z, B_p^o) < \partial]\!] \quad (13)$$

in which, $d(\ )$ stands for the Euclidean distance and $\partial$ is usually set as 0.75% of the image diagonal.

Then, for the category $o$, we can get:

$$F_1^c = \frac{2*P_o*R_o}{P_o + R_o} \quad (14)$$

To finally generate the bfscore for each image, we can average $F_1^c$ over all classes. Similarly, we can average bfscore of each image over the whole dataset to obtain the dataset's bfscore.

Intersection over union (IoU), which is also known as the Jaccard similarity coefficient, can be utilized if we want to provide a statistical accuracy measure that helps to better reveal false positives. IoU is calculated by the ratio of correctly classified pixels to the number of ground truth and predicted pixels in that category. Weighted IoU (wIoU) is mainly used to measure the performance of the model tested on a disproportionally sized classes, which aims to exclude the impact of errors in the small classes on the aggregate quality score.

$$IoU = \frac{1}{L}\sum_{a=1}^{L}\frac{P_{aa}}{M_a + H_b - P_{aa}} \quad (15)$$

However, as our dataset is severely imbalanced across categories, the mean IoU may not be an appropriate measure. Therefore, we used the wIoU instead of the mean IoU to measure the performance of the proposed algorithm.

$$wIoU = \frac{1}{\sum_{a=1}^{n}\sum_{b=1}^{n}P_{ab}}\sum_{a=1}^{n}\frac{\sum_{b=1}^{n}P_{ab}P_{aa}}{\sum_{b=1}^{n}P_{ab} + \sum_{b=1}^{n}P_{ba} - P_{aa}} \quad (16)$$

Per image score: as we need to avoid getting an algorithm which works extremely well on some images but poor on most images, it is necessary to check performance of the model not only for individual pixels in our tests set but also assess how the model works for each individual image. Second, the per image score can help to reduce the bias towards the large objects, which is because the missing segmented least objects have a small impact on the confusion matrix. Third, per image score enables drawing realistic comparisons of image segmentation results produced by different algorithms with segmentation produced by clinical experts. Therefore, in this study, we also use the per image score as a performance metric.

## 4. Experiments and Results

### *4.1 Data preparation*

The data is collected from patients with MI. To date we collected 1822 raw images which have been manually annotated by human experts. The muscle, blood area, scar, microvascular obstruction (MVO) and background were contoured by expert clinicians. The reference ground truth is then obtained as the employed experts manually label the raw images at the pixel level according to the contour information into the following five categories: background, blood, muscle, scar and MVO using Image Labeler integrated into MATLAB. In this research, the MVO and scar areas are then combined, and both considered as the scar due to the limited MVO data. The MRI images are resized to 256 by 256. Figure 11 shows an example of the raw image (Figure 11(a)) and raw image with labels (Figure 11(b)), of which the red areas represent the background, and respectively, the blue area, for blood pool, yellow areas for scar and the areas for myocardium. Unfortunately, the sample raw image does not include the MVO area as the MVO only appears in very few images. After the data preparation, we divide the whole dataset into 60% for training, 20% for the validation and the remaining 20% for the test. To avoid leakage of information from the training data into the test set, images in the tests set were taken from a cohort of patients whose MRI scans were not present in the training/validation sets.

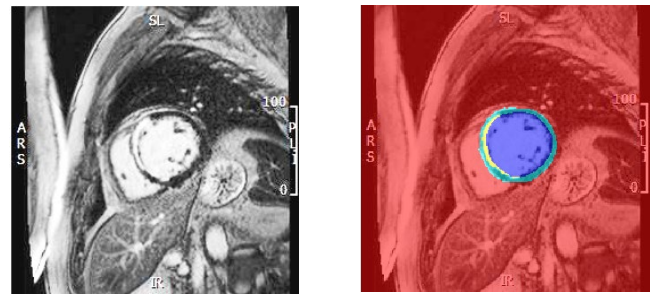

**(a)** raw MRI image      **(b)** raw image with labels

**Figure 11.** An example of the raw image and labelled image.

### *4.2 Experiment environment*

All the experiments were carried on a workstation with a 1.99 GHz processor and 16 GB memory with Windows operation system. The proposed algorithm is implemented in Matlab without optimization. The Training parameters are shown in Table 2. To make the comparison fair, we kept the training parameters same as shown in Table 2. We trained the proposed model via the Stochastic gradient descent method (SGDM). The initial learning rate was set as $e^{-3}$, the learning rate drop period was set to 10, and the learning rate drop factor was 3. The max number of epochs was 50, and the mini batch size was set to 10. All the algorithms were executed under the environment of GPU, to accelerate the computing speed. The training algorithm would stop either because the max epochs were reached or the stop criteria were met as we set the validation patience as 4.

**Table 2.** Training parameters

| Name | Parameters |
|---|---|
| Training algorithm | data |
| Learn Rate Drop Period | 10 |
| Learn Rate Drop Factors | 3 |
| Initial Learn Rate | $e^{-3}$ |
| Max epochs | 50 |
| Mini batch size | 10 |
| Execute environment | GPU |
| Validation patience | 4 |

*4.3 Segmentation result based of proposed method*

In order to explore flexibility of our method and optimise performance, we built different models based on different feature extraction methods and name the three corresponding models as: MI-ResNet50-AC, MI-ResNet18-AC and MI-MobileNet-AC for easy remember. Due to the data imbalance, the class weights were set as: 13.7678, 0.7802, 1.3923 and 0.0163 for the scar, muscle, blood and background, respectively, as is shown in Table 3.

**Table 3.** Class weight

| Type | Weight |
|---|---|
| Training algorithm | data |
| Learn Rate Drop Period | 10 |
| Learn Rate Drop Factors | 3 |
| Initial Learn Rate | $e^{-3}$ |
| Max epochs | 50 |
| Mini batch size | 10 |
| Execute environment | GPU |
| Validation patience | 4 |

**Table 4.** Performance achieved by the proposed models

| Model | Global accuracy | Mean Accuracy | wIoU | bfscore |
|---|---|---|---|---|
| MI-MobileNet-AC | 0.9569 | 0.8202 | 0.9463 | 0.5351 |
| MI-ResNet50-AC | 0.9738 | 0.8601 | 0.9647 | 0.6446 |
| MI-ResNet18-AC | 0.9679 | 0.8483 | 0.9584 | 0.5839 |

As is seen in Table 4, for the global segmentation, MI-ReNet50-AC provides the best performance with global accuracy of 0.9738, mean accuracy 0.8601, wIoU 0.9647 and bfscore 0.6446. MI-ResNet18-AC is slightly better than MI-MobileNet-AC. However, it is seen in Table 5, that for the scar tissue, MI-ResNet18-AC provides the best performance in terms of accuracy and similar performance in terms of bfscore compared with MI-ResNet50-AC. Figure 12 shows a bar chart for a clearer comparison of performance for all three proposed models.

**Table 5.** Performance for each category based on proposed models

| | Category | Scar | Blood | Muscle | Background |
|---|---|---|---|---|---|
| MI-ResNet50-AC | Accuracy | 0.6429 | 0.8402 | 0.8779 | 0.9686 |
| | bfscore | 0.4634 | 0.6837 | 0.4022 | 0.8552 |
| MI-ResNet18-AC | Accuracy | 0.7441 | 0.8255 | 0.8511 | 0.9724 |
| | bfscore | 0.4221 | 0.6226 | 0.4187 | 0.8559 |
| MI-MobileNet-AC | Accuracy | 0.4245 | 0.8809 | 0.8567 | 0.9664 |
| | bfscore | 0.3669 | 0.5996 | 0.3729 | 0.8411 |

**Table 6.** Confusion matrix based on proposed models

| | | Target class | | | |
|---|---|---|---|---|---|
| | | Scar | Blood | Muscle | Background |
| MI-ResNet50-AC Output class | Scar | 0.6429 | 0.1557 | 0.1982 | 0.0031 |
| | Blood | 0.0543 | 0.8402 | 0.0980 | 0.0074 |
| | Muscle | 0.0352 | 0.0640 | 0.8779 | 0.0229 |
| | Background | 0 | 0.0016 | 0.0291 | 0.9686 |
| MI-ResNet18-AC Output class | Scar | 0.7441 | 0.1180 | 0.1371 | 0 |
| | Blood | 0.0774 | 0.8255 | 0.0904 | 0.0068 |
| | Muscle | 0.0676 | 0.0621 | 0.8511 | 0.0193 |
| | Background | 0.0023 | 0.0023 | 0.0230 | 0.9724 |
| MI-MobileNet-AC Output class | Scar | 0.4245 | 0.3429 | 0.2326 | 0 |
| | Blood | 0.0242 | 0.8809 | 0.0870 | 0.0079 |
| | Muscle | 0.0266 | 0.0964 | 0.8567 | 0.0203 |
| | Background | 0.0011 | 0.0051 | 0.0273 | 0.9664 |

Table 6 shows the confusion matrix based on each model for a clear performance comparison. The rows stand for the predicted class and the columns stand for the true class. The correctly classified categories are shown as the diagonal cells, and the incorrectly classified observations are shown as the off-diagonal cells.

The time analysis is based on the current training dataset. For the proposed three models: MI-MobileNet-AC, MI-ReNet50-AC and MI-ResNet18-AC cost the time 24′1′′, 57′35′′ and 24′50′′ respectively.

**Table 7.** Time analysis of the proposed algorithm

| Model | Training Time cost |
|---|---|
| MI-MobileNet-AC | 24′1′′ |
| MI-ReNet50-AC | 57′35′′ |
| MI-ResNet18-AC | 24′50′′ |

Table 7 shows that computational costs of MI-RestNet50-AC are double those of the other two methods. MI-ResNet18-AC and MI-MobileNet-AC have similar computation cost. With the validation patience as 4, MI-MobileNet-AC, MI- MI-ResNet50-AC, MI- MI-ResNet18-AC stops at epoch 7, 2 and 10 respectively (all stop earlier than the maximal number of epochs we set for these experiments).

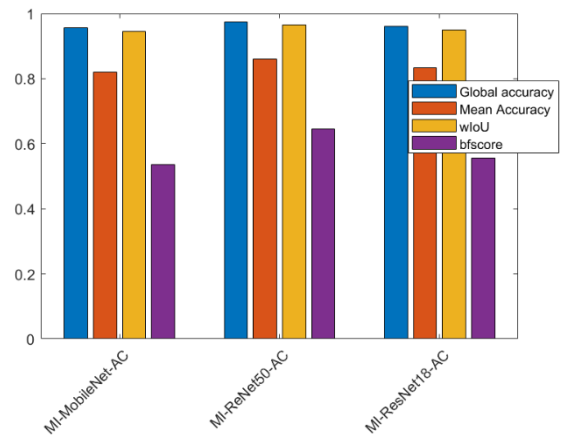

**Figure 12.** Bar chart of the performance of three proposed models.

*4.4 Segmentation result based on state of art methods*

In order to demonstrate the advantage of proposed approach and models, we compared the performance of our models to that of the state of art models, including the conventional CNN and UNet (3). A summary of performance for these models in the task of global segmentation is shown in Table 8.

**Table 8.** Comparison to the state-of-art methods

| Model | Global accuracy | Mean Accuracy | wIoU | bfscore |
|---|---|---|---|---|
| CNN | 0.6021 | 0.5632 | 0.4367 | 0.1574 |
| MI-ResNet50-AC | 0.9738 | 0.8601 | 0.9647 | 0.6446 |
| Unet | 0.6332 | 0.6222 | 0.6117 | 0.1626 |

As we can see from Table 8, our proposed model MI-ResNet50-AC provides the highest accuracy and bfscores (the Unet architecture trained on the same data provides global accuracy of 0.6332, mean accuracy 0.6222, with 0.6117 for the wIoU, and a bfscore of 0.1626). Remarkably, our network's bfscore is approximately four-fold higher than that of state-of-the art on our data.

In order to provide a clearer relation between performance of our proposed method and the most recent method Unet, we provide the scatter plot (based on per image comparison) as shown in Figure 13, Figure 13 (a), Figure 13 (b) and Figure 13 (c) show the per image global accuracy, average accuracy, bfscore comparison respectively by MI-ResNet18-AC and Unet vs the MI-ResNet50-AC. We can find that from the perspective of global segmentation for each image, MI-ResNet50-AC surpasses MI-ResNet18-AC in some cases. For a few cases, MI-ResNet18-AC performs better than MI-ResNet50-AC. However, both MI-ResNet50-AC and MI-ResNet18-AC outperformed UNet for most images in this dataset in terms of global accuracy, average accuracy and bfscore.

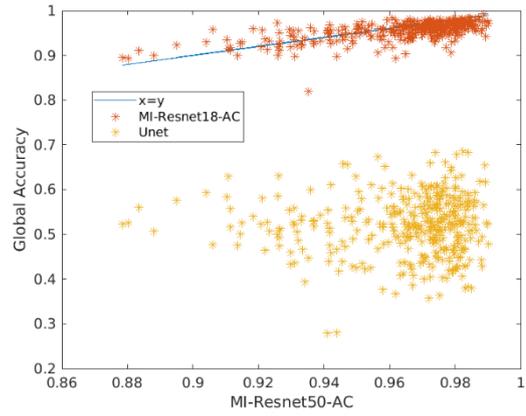

**(a)** global accuracy

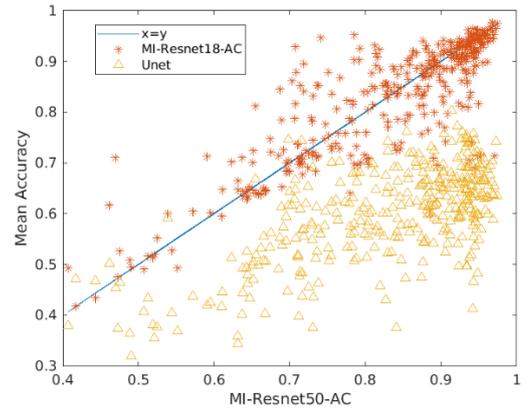

**(b)** mean accuracy

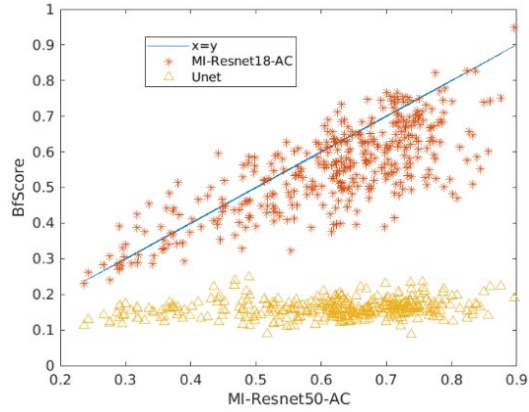

**(c)** mean accuracy

**Figure 13.** Bar Scatter plot of the performance based on per image of MI-ResNet18-AC, UNet vs MI-ResNet50-AC.

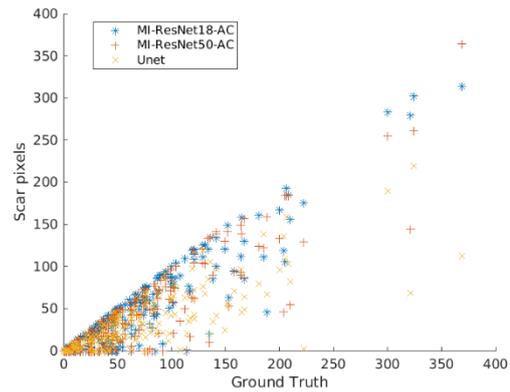

**Figure 14.** Scar detection based on MI-ResNet18-AC, MI-ResNet50-AC and UNet vs the ground truth.

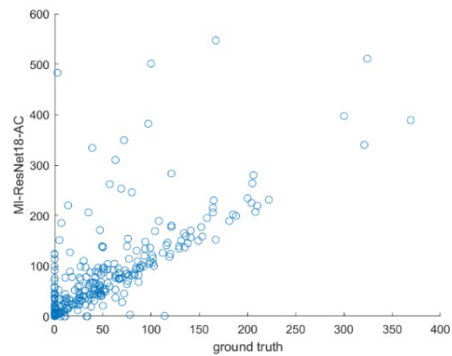

**Figure 15.** Scar detection based on MI-ResNet18-AC vs the ground truth including the false alarm.

As can be seen from the confusion matrix (shown in Table 1) and Table 5, MI-ResNet18-AC has the best performance for scar quantification compared to MI-ResNet50-AC and UNet. Figure 14 shows the scatter plot of the correctly detected scar element by MI-ResNet18-AC, MI-ResNet50-AC and UNet vs the ground truth based on per image, we can find that the MI-ResNet18-AC obviously has more cases close to the diagonal line. Figure 15 shows the scar detection including false alarms based on MI-ResNet18-AC vs the ground truth, we can find that though MI-ResNet18-AC can detect the true positives at a satisfactory rate, while the false alarm rate is still quite high for the clinical application. Therefore, in our future research, we need to pay extra attention to reduce the false alarm rate. For the clinical reference, we also show the scatter plot per case. It is apparent that further work may be needed to reduce the rate of false alarms. Error correction approaches may potentially be used to address this issue (4-6). Detailed exploration of such functionality, however, is outside of the scope of the current work.

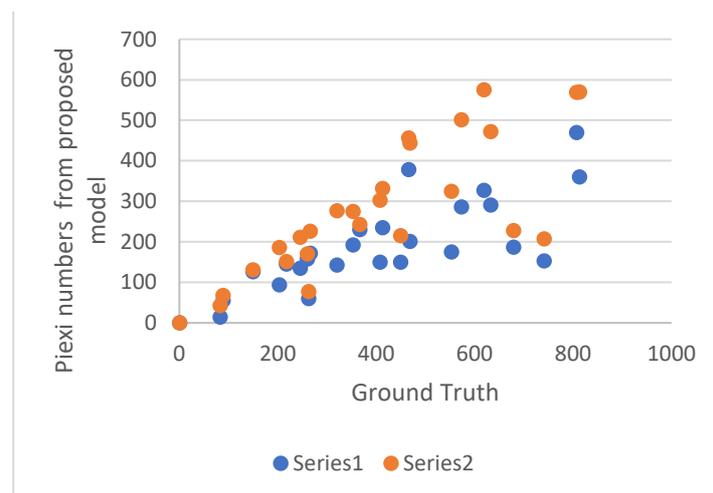

**Figure 16.** Detection result per case. Series 1 includes the false alarm and series 2 only contains the true positives.

## 5. Conclusion

In this paper, we proposed a new end-to-end method for automatic MI segmentation, as the detection and quantification of the MI is crucial for determining clinical management and prognosis. Although LGE-CMR (7) is the non-invasive "gold standard" method as it permits optimal differentiation between normal and damaged myocardium with the use of gadolinium based contrast agents and special magnetic resonance pulse sequences, to date, there is no fully automatic "Gold standard" method for the detection and quantification of MI. In this work we make a step forward towards achieving this aim with the hope to reduce the uncertainty brought by the technical variability and inherent bias of the data and labels. We propose a novel deep learning model, MyI-Net, which accommodates MI-ResNet, MI-MobileNet models as initial feature extractors and is equipped with ASPP with an add-on module for the recovery of the spatial information to compute the final segmentation output. Considering the limited size of the dataset, a data augmentation pre-processing step was integrated in our model construction pipeline. It is apparent from Figure 9 that our dataset was severely imbalanced as it contained primarily background elements, followed, in descending order, by blood pool, muscle and scar. A weight matrix was used in to minimize our classifier's bias towards any specific category. Performance of the algorithm is shown in Table 4, with the best performance for global segmentation being provided by MI-ResNet50-AC with a global accuracy as 0.9738, mean accuracy 0.8601, wIoU 0.9647, and bfscore 0.6446. In comparison with other state of art methods (Table 8), our model outperformed state-of-the-art architectures on the dataset we had access to. However, considering the detection of scar tissue, we found that MI-ResNet18-AC, being a smaller model than MI-ResNet50-AC, provides the highest scar detection accuracy and a bfscore similar to that of MI-ResNet50-AC. Furthermore, we compared the computation cost for the three proposed models. Table 7 shows that MI-ResNet50-AC required the largest amount of time as compared to the other two models, mainly because MI-ResNet50-AC is a deeper network. Considering the above summary, we integrated both MI-ResNet50-AC and MI-ResNet18-AC into our proposed system MyI-Net. In general, however, the choice of specific feature extractor model depends on a given task (e.g. MI scar detection, or global segmentation). However, one limitation of the work is the relatively small size of the dataset. To be able to build more accurate, higher-performing models, larger datasets may be required. These datasets are also necessary for testing the system at the level of individual patients. To improve generalization capabilities of the model, further data should be collected from a greater variety of scanner vendors; it should also include more data with MVO. Meanwhile, and based on the per image analysis, we can observe that the proposed model is not as stable as might be desired. Therefore, the algorithm needs to be further tuned to achieve more robust and stable performance. We are considering a fusion technique in the future to fully explore the potential of the proposed deep learning models. We also consider error correction approaches (5,6,8) in our future research to further improve the performance of our proposed method to minimize false positives.

**Author Contributions:** Conceptualization, Gerry P McCann and Ivan Y Tyukin; methodology, Shuihua Wang and Ivan Y Tyukin; software, Shuihua Wang, Ivan Y Tyukin; validation, Shuihua Wang, Ivan Y Tyukin and J Ranjit Arnold; formal analysis, Shuihua Wang, Ivan Y Tyukin, Gerry P McCann; investigation, Kelly Parke; resources, Ahmed M.S.E.K Abdelaty; data curation, Ahmed M.S.E.K Abdelaty, Kelly Parke; writing—original draft preparation, Shuihua Wang; writing—review and editing, Ivan Y Tyukin; supervision, Gerry P McCann, Ivan Y Tyukin; project administration, Gerry P McCann, Ivan Y Tyukin; funding acquisition, Gerry P McCann, Ivan Y Tyukin. All authors have read and agreed to the published version of the manuscript.

**Funding:** This research was funded by Leicester Drug Discovery & Diagnostics (LD3), Leicester 10x10 Challenge Fund and British Heart foundation Accelerator Award AA/18/3/34220. IYT is supported by the UKRI Turing AI Fellowship EP/V025295/1 and by the Ministry of Science and Higher Education of the Russian Federation (Project No. 075-15-2021-634), GPM and JRA are supported by NIHR research professorship (RP-2017-08-ST2-007) and Clinician Scientist Award (CS-2018-18-ST2-007) respectively.

**Institutional Review Board Statement:** Not applicable

**Data Availability Statement:** Not applicable

**Acknowledgments:** Not applicable

**Conflicts of Interest:** The authors declare no conflict of interest.

## References

1.  Tambe, V.; Dogra, M.; Shah, B. S.; and Hegazy, H. TRAUMATIC RCA DISSECTION AS A CAUSE OF INFERIOR WALL ST ELEVATION MI. *Chest*, vol. 154, **2018**, pp. 87A-88A.
2.  Jia, S.; Weng, L.; and Zheng, M. Nogo-C causes post-MI arrhythmia through increasing calcium leakage from sarcoplasmic reticulum. *Journal of Molecular and Cellular Cardiology*, vol. 140, **2020**, p. 43.
3.  Heart Statistics. Available online: https://www.bhf.org.uk/what-we-do/our-research/heart-statistics (accessed on 25/Oct/2022).
4.  Fischer, K; Obrist, S.J.; Erne, S.A.; Stark, A.W.; Marggraf, M.; Kaneko, K. et al. Feature Tracking Myocardial Strain Incrementally Improves Prognostication in Myocarditis Beyond Traditional CMR Imaging Features. *JACC: Cardiovascular Imaging*, vol. 13, , **2020**, pp. 1891-1901.
5.  Kuetting, D.L.R.; Homsi, R.; Sprinkart A.M. et al. Quantitative assessment of systolic and diastolic function in patients with LGE negative systemic amyloidosis using CMR. *International Journal of Cardiology*, **2017**, 232, pp. 336-341.
6.  Kurt, M.; Yurtseven, H.; and Kurt, A. Calculation of the Raman and IR frequencies as order parameters and the damping constant (FWHM) close to phase transitions in methylhydrazinium structures. *Journal of Molecular Structure*, **2019**, vol. 1181, pp. 488-492.
7.  Hien, N.D. Comparison of magneto-resonance absorption FWHM for the intrasubband/intersubband transition in quantum wells. *Superlattices and Microstructures*, **2019**, vol. 131, pp. 86-94.
8.  Eitel, I.; Desch, S.; de Waha, S.; Fuernau, G.; Gutberlet, M.; Schuler, G. et al., Long-term prognostic value of myocardial salvage assessed by cardiovascular magnetic resonance in acute reperfused myocardial infarction. *Heart*, **2011**, vol. 97, pp. 2038-2045.
9.  Amado, L.C.; Gerber, B.L.; Gupta, S.N.; Rettmann, D.W.; Szarf, G.; Schock, R. et al., Accurate and objective infarct sizing by contrast-enhanced magnetic resonance imaging in a canine myocardial infarction model. *J Am Coll Cardiol*, **2004**, vol. 44, pp. 2383-2389.
10. Flett, A. S.; Hasleton, J.; Cook, C.; Hausenloy, D.; Quarta, G.; Ariti, C. et al., Evaluation of techniques for the quantification of myocardial scar of differing etiology using cardiac magnetic resonance. *JACC Cardiovasc Imaging*, **2011**, vol. 4, pp. 150-156.
11. Hsu, L.Y.; Natanzon, A.; Kellman, P.; Hirsch, G.A.; Aletras, A.H.; Arai, A.E. Quantitative myocardial infarction on delayed enhancement MRI. Part I: Animal validation of an automated feature analysis and combined thresholding infarct sizing algorithm. *J Magn Reson Imaging*, **2006**, vol. 23, pp. 298-308.
12. Tong, Q.; Li, C.; Si, W.; Liao, X.; Tong, Y.; Yuan, Z. et al., RIANet: Recurrent interleaved attention network for cardiac MRI segmentation. *Computers in Biology and Medicine*, **2019**, vol. 109, pp. 290-302.
13. Shan, F.; Gao, Y.; Wang, J.; Shi, W.; Shi, N.; Han, M. et al., Lung Infection Quantification of COVID-19 in CT Images with Deep Learning. arXiv preprint arXiv:2003.04655, **2020**
14. Xu, C.; Xu, L.; Gao, Z.; Zhao, S.; Zhang, H.; Zhang, Y. et al., Direct Detection of Pixel-Level Myocardial Infarction Areas via a Deep-Learning Algorithm. Presented at the *Medical Image Computing and Computer Assisted Intervention* (MICCAI), **2017**, pp. 240-249.
15. Bleton, H.; Margeta, J.; Lombaert, H.; Delingette , H.; Ayache, N. Myocardial Infarct Localization Using Neighbourhood Approximation Forests. Presented at the *Statistical Atlases and Computational Models of the Heart* (STACOM 2015), **2015**, pp. 108-116.
16. Fahmy, A.S.; Rausch, J.; Neisius, U.; Chan, R.H.; Maron, M.S.; Appelbaum, E. et al., Automated Cardiac MR Scar Quantification in Hypertrophic Cardiomyopathy Using Deep Convolutional Neural Networks. *JACC: Cardiovascular Imaging*, **2018**, vol. 11, pp. 1917-1918.
17. Bernard, O.; Lalande, A.; Zotti, C.; Cervenansky, F.; Yang, X.; Heng, P. et al., Deep Learning Techniques for Automatic MRI Cardiac Multi-Structures Segmentation and Diagnosis: Is the Problem Solved?. *IEEE Transactions on Medical Imaging*, **2018**, vol. 37, pp. 2514-2525.
18. Ronneberger, O.; Fischer, P.; Brox, T. U-Net: Convolutional Networks for Biomedical Image Segmentation. Springer, *Cham*, **2015**, pp.234-241.
19. Fahmy, A.S.; Neisius, U.; Chan, R.H.; Rowin, E.J.; Manning, W.J.; Maron, M.S. et al., Three-dimensional Deep Convolutional Neural Networks for Automated Myocardial Scar Quantification in Hypertrophic Cardiomyopathy: A Multicenter Multivendor Study. *Radiology*, **2020**, vol. 294, pp. 52-60.
20. New Findings Confirm Predictions on Physician Shortage. Available online: https://www.aamc.org/news-insights/press-releases/new-findings-confirm-predictions-physician-shortage (accessed on 25/Oct/2022).
21. Clinical radiology UK workforce census report 2018. Available online: https://www.rcr.ac.uk/publication/clinical-radiology-uk-workforce-census-report-2018 (accessed on 25/Oct/2022).
22. Haque, I.R.I.; Neubert, J. Deep learning approaches to biomedical image segmentation. *Informatics in Medicine Unlocked*, **2020**, vol. 18, p. 100297.
23. Luo, L.; Yang, Z.; Cao, M.; Wang, L.; Zhang, Y.; Lin, H. A neural network-based joint learning approach for biomedical entity and relation extraction from biomedical literature. *Journal of Biomedical Informatics*, **2020**, vol. 103, p. 103384.

24. Moradi, M.; Dorffner, G.; Samwald, M. Deep contextualized embeddings for quantifying the informative content in biomedical text summarization. *Computer Methods and Programs in Biomedicine*, **2020**, vol. 184, p. 105117.
25. Nam, J.G.; Park, S.; Hwang, E.J.; Lee, J.H.; Jin, K.N.; Lim, K.Y. et al., Development and Validation of Deep Learning–based Automatic Detection Algorithm for Malignant Pulmonary Nodules on Chest Radiographs. *Radiology*, **2019**, vol. 290, pp. 218-228.
26. Wang, S.H.; G. McCann, G.; Tyukin, I. Myocardial Infarction Detection and Quantification Based on a Convolution Neural Network with Online Error Correction Capabilities. In 2020 *International Joint Conference on Neural Networks* (IJCNN), **2020**, pp. 1-8.
27. Krizhevsky, A.; Sutskever, I.; and Hinton, G.E. ImageNet Classification with Deep Convolutional Neural Networks. *Communications of the ACM*, **2017**, 60(6), pp. 84-90.
28. Tyukin, I.Y.; Gorban, A.N.; Sofeykov, K.I.; Romanenko, I. Knowledge Transfer Between Artificial Intelligence Systems. *Frontiers in Neurorobotics*, **2018**, 12, p. 49.
29. He, K.; Zhang, X.; Ren, S.; Sun, J. Deep Residual Learning for Image Recognition. in *2016 IEEE Conference on Computer Vision and Pattern Recognition* (CVPR), **2016**, pp. 770-778.
30. Sandler, M.; Howard, A.; Zhu, M.; Zhmoginov, A.; Chen, L. MobileNetV2: Inverted Residuals and Linear Bottlenecks. In *2018 IEEE/CVF Conference on Computer Vision and Pattern Recognition*, **2018**, pp. 4510-4520.
31. Howard, A.G.; Zhu, M.; Chen, B.; Kalenichenko, D.; Wang, W.; Weyand, T. et al., Mobilenets: Efficient convolutional neural networks for mobile vision applications. arXiv preprint arXiv:1704.04861, **2017**.
32. Li, H.; Qi, F.; Shi, G.; Lin, C. A multiscale dilated dense convolutional network for saliency prediction with instance-level attention competition, *Journal of Visual Communication and Image Representation*. **2019**, vol. 64, p. 102611
33. Friston, K.J.; Rosch, R.; Parr, T.; Price, C.; Bowman, H. Deep temporal models and active inference. *Neuroscience & Biobehavioral Reviews*, **2018**, vol. 77, pp. 388-402.
34. Tyukin, I.Y.; Gorban, A.N.; Green, S.; Prokhorov, D. Fast construction of correcting ensembles for legacy Artificial Intelligence systems: Algorithms and a case study. *Information Sciences*, **2019**, vol. 485, pp. 230-247.
35. Gorban, A.N.; Golubkov, A.; Grechuk, B.; Mirkes, E.M.; Tyukin, I.Y. Correction of AI systems by linear discriminants: Probabilistic foundations. *Information Sciences*, **2018**, vol. 466, pp. 303-322.
36. Yoneyama, M.; Zhang, S.; Hu, H.H.; Chong, L.R.; Bardo, D.; Miller, J.H. et al., Free-breathing non-contrast-enhanced flow-independent MR angiography using magnetization-prepared 3D non-balanced dual-echo Dixon method: A feasibility study at 3 Tesla. *Magnetic Resonance Imaging*, **2019**, vol. 63, pp. 137-146.
37. Poliak, M.; Tomicová, J.; Jaśkiewicz, M. Identification the Risks Associated With the Neutralization of the CMR Consignment Note. *Transportation Research Procedia*, **2020**, vol. 44, pp. 23-29.
38. Gorban, A.N.; Burton, R.; Romanenko, I.; Tyukin, I.Y. One-trial correction of legacy AI systems and stochastic separation theorems. *Information Sciences*, **2019**, vol. 484, pp. 237-254.